# On the Origin of LLMs:

## An Evolutionary Tree and Graph for 15,821 Large Language Models

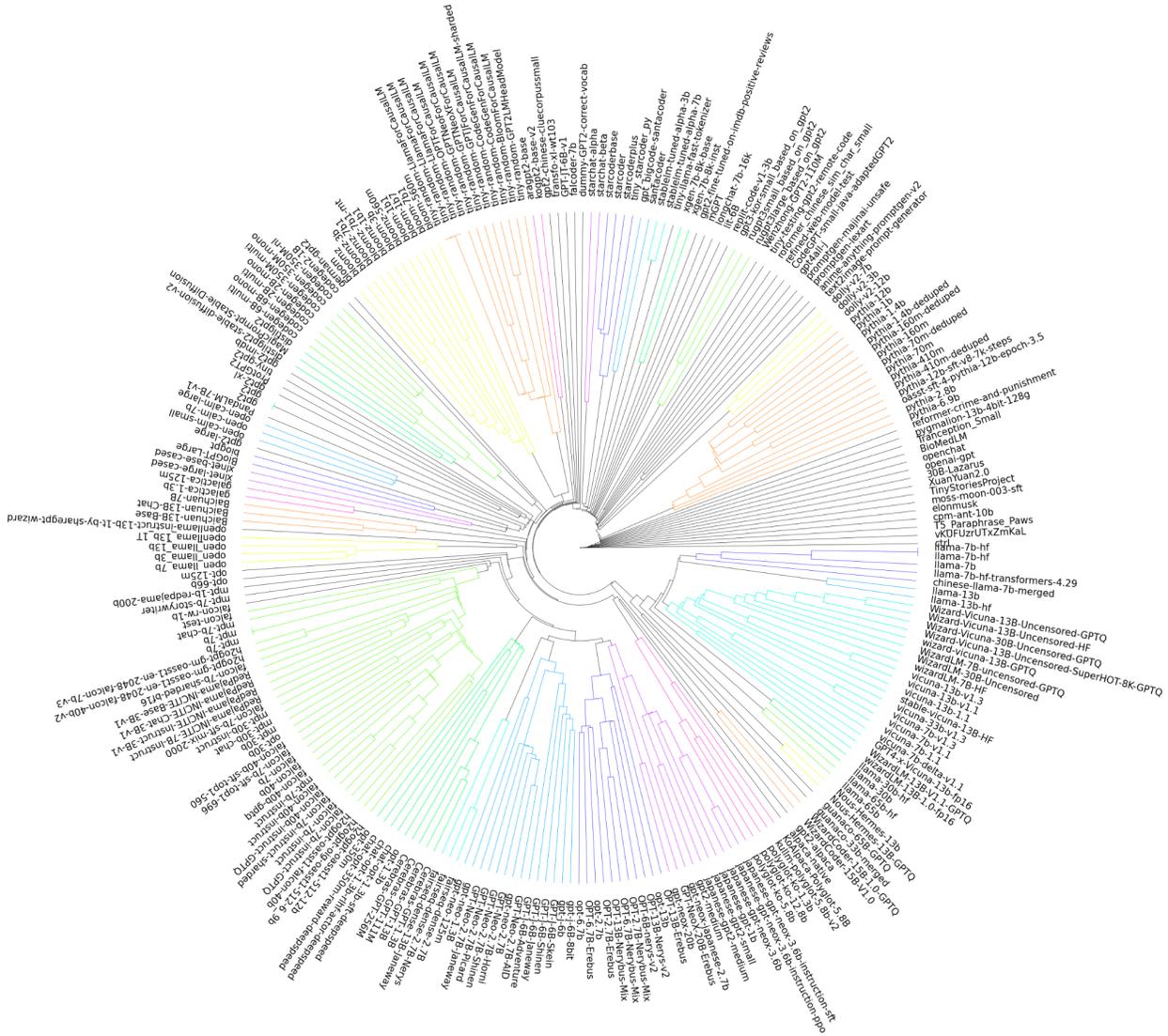

Sarah R Gao, Andrew K Gao

Canyon Crest Academy, Stanford University

# Abstract

Since late 2022, Large Language Models (LLMs) have become very prominent with LLMs like ChatGPT and Bard receiving millions of users. Hundreds of new LLMs are announced each week, many of which are deposited to Hugging Face, a repository of machine learning models and datasets. To date, nearly 16,000 Text Generation models have been uploaded to the site. Given the huge influx of LLMs, it is of interest to know which LLM backbones, settings, training methods, and families are popular or trending. However, there is no comprehensive index of LLMs available. We take advantage of the relatively systematic nomenclature of Hugging Face LLMs to perform hierarchical clustering and identify communities amongst LLMs using n-grams and term frequency-inverse document frequency. Our methods successfully identify families of LLMs and accurately cluster LLMs into meaningful subgroups. We present a public web application to navigate and explore Constellation, our atlas of 15,821 LLMs. Constellation rapidly generates a variety of visualizations, namely dendrograms, graphs, word clouds, and scatter plots. Constellation is available at the following link: https://constellation.sites.stanford.edu/.

The dataset we created will be shared publicly on Github, under @andrewgcodes (https://github.com/andrewgcodes).

# Introduction

Large language models (LLMs) are trained to generate realistic text given a user prompt [1]. Popular LLMs include ChatGPT, Bard, and the LLaMa family of models [2]. In addition to large companies like OpenAI and Google, smaller research groups and individuals can also train LLMs and share them through Hugging Face, a popular machine learning repository [3,4]. As of July 18, 2023 at 12 PM (GMT -5), 15,821 LLMs (or at least, Text Generation models) were available publicly on Hugging Face. To our knowledge, few attempts have been made to organize these LLMs, perhaps due to the immense number of models. Inspired by the bioinformatics technique of using hierarchical clustering on DNA sequences, we apply hierarchical clustering to the Hugging Face model names, assuming that similar names indicate similarity [5]. We also construct a graph of LLMs and detect communities using the Louvain method. Additionally, we generate other visualizations and explore the data.

# Methods

## Libraries

- BeautifulSoup [6]
- Pandas [7]
- Streamlit [8]
- Scipy [9]
- Plotly [10]
- Numpy [11]
- Scikit-learn [12]
- Radial Tree [13]
- NLTK [14]
- Matplotlib [15]
- Python-Louvain [16]
- NetworkX [17]
- Wordcloud [18]
- RegEx [19]

## Data Collection

Python's BeautifulSoup library was used to retrieve the names, number of likes, and number of downloads of Hugging Face models labeled with "Text Generation". Data collected included model names, Readme links, number of downloads, and the number of likes. Data collection was performed on July 18, 2023 around 12 PM (US ET; GMT -5). Note that data collection was not instantaneous. In some instances, we failed to retrieve a number of likes or downloads for a model.

## Parameter Extraction

In addition to the above attributes, model parameters were inferred from the model name using a regular expression (RegEx) pattern (\d+(\.\d+)?)(B|M|b|m). This pattern matches digit sequences followed by "B", "M", "b", or "m", as model sizes are often included in the name (e.g., "falcon-7b"). The number of parameters in millions was recorded in a column named 'params_millions' in the dataset. If parameters couldn't be inferred, the corresponding field was marked as 'NaN'.

## Data Analysis and Visualization

We used libraries such as Scipy, Plotly, Numpy, Scikit-learn, Radial Tree, NLTK, and Matplotlib to analyze the dataset and generate visualizations. A full list of imports is provided at the beginning of Methods.

## Text Feature Extraction:

The model names were converted into a matrix of Term Frequency-Inverse Document Frequency (TF-IDF) features using Scikit-learn's TfidfVectorizer. The vectorizer was configured to break down the model names into n-grams ranging from 2 to 8 characters.

## Hierarchical Clustering:

Hierarchical clustering with single linkage was performed using the matrix of TF-IDF features. Cosine distance was used as a similarity measure between the model names. The clustering result was visualized as an interactive dendrogram using the Plotly library.

## Agglomerative Clustering:

In addition to hierarchical clustering, agglomerative clustering was also performed with a specified number of clusters. The size of each cluster was calculated and visualized as a bar chart.

## Word Clouds

To understand the contents of our agglomerative clusters, we generate Word Clouds the most common n-grams in each cluster, with more frequent n-grams being presented with larger text size.

## Graph Visualization with Communities

A graph-based visualization was constructed to provide an intuitive understanding of the relationship and similarity among the models. NetworkX, a Python library for the creation, manipulation, and study of the structure, dynamics, and functions of complex networks, was used to generate the graph.

### Node Creation:

Each model name was represented as a node in the graph. The graph was initialized as an undirected graph, and each model name was added as a node using the 'add_node' function. The model name served as the node label.

### Edge Creation:

Edges in the graph were used to represent the similarity between pairs of model names. After calculating the cosine similarity matrix, an edge was added between two nodes (model names) if their cosine similarity was above a specific threshold (0.2 in this case). The cosine similarity value was set as the weight of the edge.

### Community Detection:

The Louvain method, a popular community detection algorithm, was used to find communities within the constructed graph. Communities represent groups of models that are more similar to each other than to models in other groups. The detected communities were used for subsequent visual enhancements.

### Layout Calculation:

The Fruchterman-Reingold force-directed algorithm was employed to calculate the layout of nodes in the graph. This algorithm arranges the nodes in such a way that all the edges are more or less equally long and there are as few crossing edges as possible.

### Interactive Visualization:

The generated graph was visualized interactively using the Plotly library. Each node represented a model and was color-coded based on the community it belonged to. Edges between the nodes indicated similarity, with their thickness corresponding to the cosine similarity score. Hovering over the nodes displayed more details about the model.

### Additional Enhancements:

The centroid (center point) of each community was computed to add a colored background for each community cluster. The size of the background color patch represented the size of the community.

## Web Application

We built a public web application using the Streamlit framework to generate interactive dendrograms, word clouds, and graphs for the data, which is available here: https://constellation.sites.stanford.edu/.

# Results

There were 15,821 public models labeled with Text Generation on Hugging Face at the time of data collection. We assembled a final Pandas dataframe containing seven columns: rank,

model_name, link, downloads, likes, ReadMeLink, and params_millions. Rank is assigned in order of number of downloads. For instance, "gpt2" has the most downloads. Note that "gpt2" does not have an inferred number of parameters because the model name does not contain any evidence of parameter size. We were able to infer model parameters for 4,560 models (28.8%). We expect our RegEx expression to result in few false positives. Not all links in ReadMeLink lead to a valid Readme file. The links were automatically computed by appending "/raw/main/README.md" to the model link. All model links should lead to a valid Hugging Face page.

| rank | model_name | link | downloads | likes | ReadMeLink | params_millions |
|---|---|---|---|---|---|---|
| 1 | gpt2 | https://huggingface.co/gpt2 | 13600000.0 | 1260.0 | https://huggingface.co/gpt2/raw/main/README.md | NaN |
| 2 | mpt-7b-instruct | https://huggingface.co/mosaicml/mpt-7b-instruct | 3050000.0 | 418.0 | https://huggingface.co/mosaicml/mpt-7b-instruct/raw/main/README.md | 7000.0 |
| 3 | bloom-560m | https://huggingface.co/bigscience/bloom-560m | 1520000.0 | 224.0 | https://huggingface.co/bigscience/bloom-560m/raw/main/README.md | 560.0 |
| 4 | distilgpt2 | https://huggingface.co/distilgpt2 | 1140000.0 | 227.0 | https://huggingface.co/distilgpt2/raw/main/README.md | NaN |
| 5 | vicuna-7b-v1.1 | https://huggingface.co/lmsys/vicuna-7b-v1.1 | 1020000.0 | 60.0 | https://huggingface.co/lmsys/vicuna-7b-v1.1/raw/main/README.md | 7000.0 |

*Figure 1.* First five rows of our dataset in order of number of downloads.

We computed a Pearson correlation coefficient of 0.242 between the number of likes and downloads a model receives. There is a clear positive but weak relationship. It is possible that this weakness indicates a disparity between model usefulness and popularity. Alternatively, larger, more powerful models may attract more attention (receiving more likes) but will garner relatively few downloads because they are too large for most Hugging Face users to use. In general, models tend to receive far more downloads than likes. This could be because there is no benefit to the user to like a model on Hugging Face, while downloading the model is beneficial. We generate a radial dendrogram on all models with over 5,000 downloads to compactly visualize relationships and families. From the dendrogram, families of LLMs like Wizard, Pythia, CausalLM, and Bloom can be observed. We suggest using the web application to view the dendrogram since the large number of leaves makes it difficult to render on a single static image clearly.

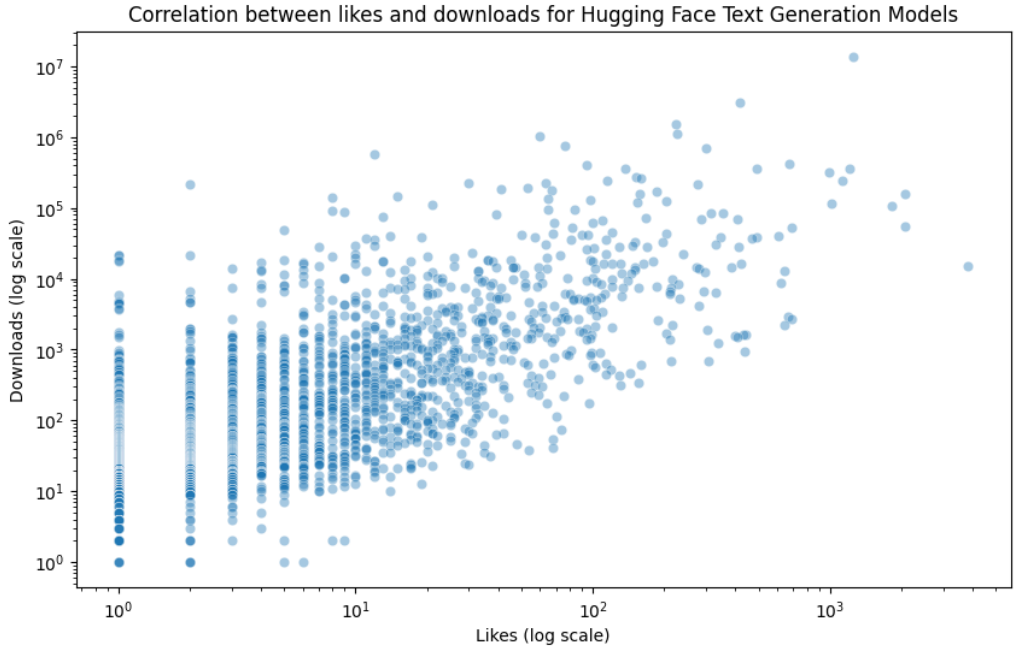

*Figure 2.* Scatter plot showing the relationship between the number of likes and downloads a model receives. Both axes received a log scale transformation.

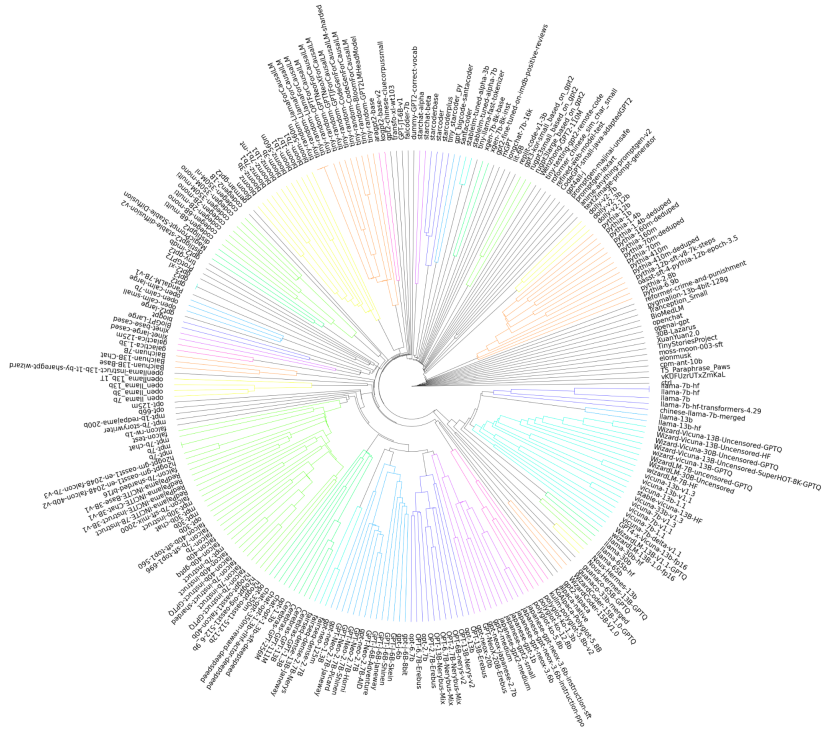

*Figure 3.* Radial dendrogram of models with over 5,000 downloads. High resolution image available on Constellation web site.

We do not show all the word clouds here due to space, but here are some of the example word clouds for clusters generated from all models with over 1,000 downloads (clusters = 20). The word clouds are helpful in understanding which model families are prominent.

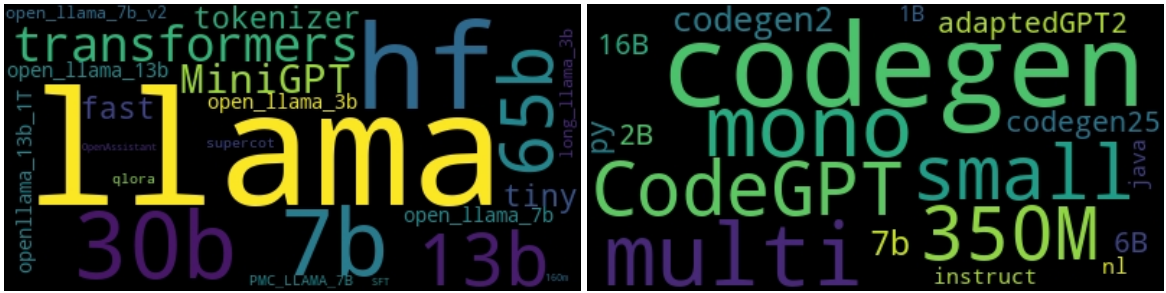

*Figure 4.* Word clouds show clusterization of LLaMa models and code-specific LLMs.

We generate a graph of the models, with similar models receiving an edge. We use the Louvain method to detect communities.

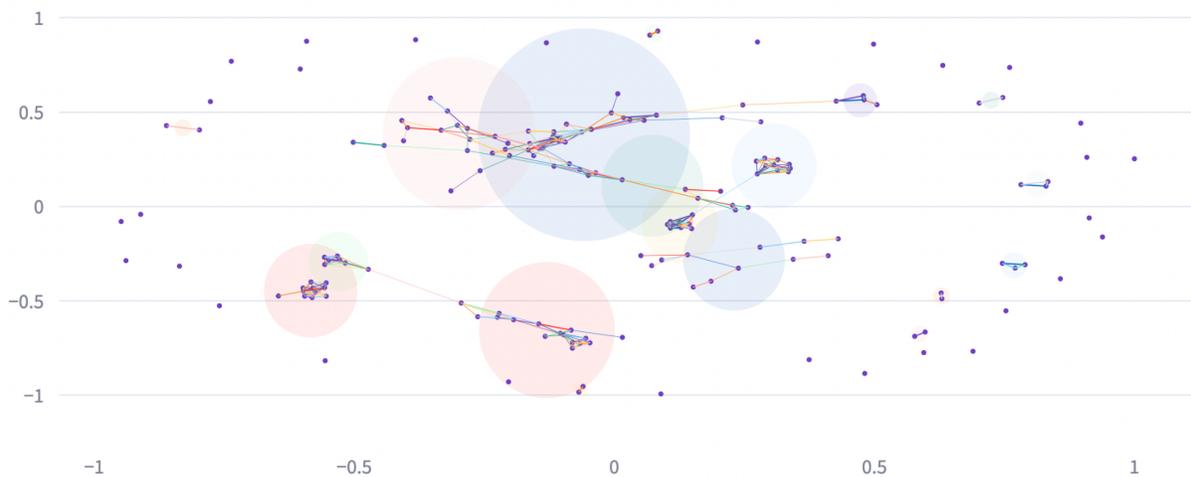

*Figure 5.* Graph of models with more than 10,000 downloads, with communities detected using the Louvain method.

We present a publicly available web application (https://constellation.sites.stanford.edu/) to dynamically explore the data. The web application enables the user to specify the minimum number of downloads an LLM must have to be considered in the analysis. The web app quickly generates a dendrogram, word clouds, and graph. Hovering over graph nodes reveals additional metadata about the model located at that node. The web application also displays useful statistics and an interactive scatter plot of likes versus downloads. Hovering over points reveals the model name.

Enter the minimum number of downloads an LLM must have to be considered.

10000                                                                                    −    +

Number of clusters to group into.

20                                                                                       −    +

☑ Show word clouds?

Run Clustering

*Figure 6.* Screenshot of the web application.

# Conclusion

The increasing number and diversity of Large Language Models (LLMs) necessitate a comprehensive and systematic approach to organize, classify, and understand these models. In this study, we have proposed an effective solution by creating Constellation, a user-friendly web application that visualizes the hierarchical relationships among LLMs, helping to reveal prominent LLM families and underlying structures.

Our approach is generally inspired by bioinformatics and sequence similarity. It utilizes hierarchical and agglomerative clustering, combined with an array of techniques such as dendrograms, word clouds, and graph-based representations. The word clouds provide a high-level view of prominent model families, while the graph-based visualization depicts the relationships and similarities among models in a more flexible format than the dendrogram.

The major limitation of our study is that it assumes that LLMs are only similar if they have similar names. This is not completely true: LLMs can be named anything by the creator who deposits it to Hugging Face. However, in general, we note that LLMs tend to be named in a structured, logical fashion. Our results indicate that our assumption that in general similar LLMs share similar names is sound. We acknowledge that our approach can miss similar LLMs, especially if one of the LLMs is arbitrarily named. Another limitation is that not all models labeled "Text Generation" are necessarily LLMs. Finally, a further caveat is that the dendrogram is not a true "evolutionary" tree. While models in the same low-level cluster are generally reliably related, this does not hold for higher-level clusters.

By making Constellation publicly available, we hope to encourage more systematic and informed engagement with LLMs. As the landscape of LLMs continues to evolve rapidly, tools such as Constellation will be instrumental in assisting the researcher and developer communities in keeping pace with these developments.

# Appendix

| Word | Occurrences |
| --- | --- |
| gpt2 | 1597 |
| 7b | 889 |
| 13b | 770 |
| gpt | 756 |
| finetuned | 611 |
| llama | 475 |
| gptq | 393 |
| distilgpt2 | 383 |
| pythia | 381 |
| model | 309 |
| wikitext2 | 297 |
| small | 294 |
| base | 285 |
| instruct | 262 |
| neo | 261 |
| opt | 252 |
| vicuna | 238 |
| 4bit | 224 |
| bloom | 215 |
| v2 | 214 |
| 30b | 203 |
| 6b | 191 |
| alpaca | 190 |
| 125m | 182 |
| codeparrot | 178 |
| rarity | 172 |
| v1 | 171 |
| falcon | 168 |
| 8k | 167 |
| sft | 167 |
| large | 166 |
| dialogpt | 160 |
| test | 157 |
| 2 | 156 |
| all | 155 |
| medium | 154 |
| lora | 153 |
| ds | 146 |
| merged | 145 |
| superhot | 143 |
| j | 141 |
| hf | 141 |
| fp16 | 133 |
| chat | 129 |
| open | 126 |
| concat | 126 |
| owt2 | 126 |
| 350m | 123 |
| 70m | 123 |
| chinese | 122 |
| 128g | 121 |
| mpt | 118 |
| gpt4 | 118 |
| 3b | 116 |
| myawesomeeli5c lm | 108 |
| tiny | 106 |
| 1 | 106 |
| 4 | 105 |
| 8bit | 103 |
| headless | 101 |
| codegen | 97 |
| 33b | 97 |
| deduped | 95 |
| sharded | 89 |
| airoboros | 88 |
| finetunedgpt2 | 85 |
| v3 | 83 |
| wizardlm | 83 |
| generator | 83 |
| no | 82 |
| 560m | 81 |
| random | 79 |
| uncensored | 74 |
| finetune | 74 |
| xl | 72 |
| 1b | 71 |
| mod | 71 |
| 27b | 69 |
| 65b | 68 |
| elonmusk | 68 |
| bloomz | 66 |
| v0 | 66 |
| bert | 65 |
| guanaco | 64 |
| ft | 64 |
| imdb | 63 |
| gptj | 62 |
| 160m | 61 |
| gptneo | 61 |
| redpajama | 58 |
| neox | 57 |

| ggml | 57 |
|------|-----|
| the | 55 |
| 5 | 55 |
| dolly | 53 |
| 12b | 53 |
| cut | 53 |
| guten | 53 |
| 67b | 52 |
| delta | 52 |

*Table 1.* Most common words/phrases among all Hugging Face LLMs